\documentclass[aps, prl, twocolumn, superscriptaddress]{revtex4}

\usepackage{amsfonts}           
\usepackage{amssymb}            
\usepackage{amsmath}            
\usepackage{comment}
\usepackage{latexsym}           
\usepackage{dcolumn}            
\usepackage[dvips]{graphicx}   
\usepackage{enumerate}
\usepackage{epstopdf}           
\usepackage{fancyhdr}           
\usepackage{hyperref}           
\usepackage{setspace}           
\usepackage{xspace}             
\usepackage{subfigure}          
\usepackage{color}   
\usepackage{bm}                 
\usepackage[ulem=normalem]{changes}

\newcommand{\rw}{\rightarrow}
\newcommand{\bea}{\begin{eqnarray}}
\newcommand{\eea}{\end{eqnarray}}

\newcommand{\cf}{{\it cf.}~}

\newcommand{\eps}{\epsilon}

\newcommand{\eg}{{\it eg.~}}
\newcommand{\ie}{{\it ie.\,}}

\newcommand{\cP}{{\cal P}}

\begin{document}

\title{The Universality Class of Nano-Crystal Plasticity: Self-Organization and Localization in Discrete Dislocation Dynamics}
\date{\today}

\author{Hengxu Song}
\affiliation{Department of Mechanical and Aerospace Engineering, West Virginia University, Morgantown, WV26506, United States.}
\affiliation{Department of Mechanical Engineering, Johns Hopkins University, Baltimore, MD21218, United States.}
\author{Dennis Dimiduk}
\affiliation{Department of Materials Science Engineering, Ohio State University, Columbus, OH43210, United States.}
\author{Stefanos Papanikolaou$^*$}
\affiliation{Department of Mechanical and Aerospace Engineering, West Virginia University, Morgantown, WV26506, United States.}
\affiliation{Department of Mechanical Engineering, Johns Hopkins University, Baltimore, MD21218, United States.}


\begin{abstract}
The universality class of the avalanche behavior in nanocrystals under uniaxial compression has been typically described as being statistically similar to the plastic response of amorphous solids, polycrystals, frictional contacts and earthquakes, despite the vast differences of the microscopic plasticity defects' character. A characteristic outcome of the crystal dislocations' character is that in macroscopic crystals under uniaxial compression, the  flow stress is known to dramatically increase at high loading rates. We investigate the effect of loading rate by performing simulations of a two-dimensional discrete dislocation dynamics model that minimally captures the  phenomenology of nanocrystalline deformation. In the context of this model, we demonstrate that a classic rate-dependence of dislocation plasticity at large rates ($>10^3/s$), fundamentally controls the system's statistical character as it competes with dislocation nucleation:
At small rates, plasticity localization dominates in small volumes. At large rates, the behavior is statistically dominated by long-range correlations of ``dragged" mobile dislocations. The resulting behavior suggests that the experimentally relevant quasi-static loading limit of crystal plasticity in small volumes belongs in a unique universality class that is characterized by a spatial integration of avalanche behaviors at various distances from a parent critical point.  
\end{abstract}
\maketitle 
{
Crystal plasticity in small volumes has been investigated in the last two decades through the compression of micro and nanopillars~\cite{Greer:2011kq, uchic2009plasticity,greer2005size,greer2006nanoscale,Ryu:2015aa,Krebs:2017aa}. In these small volumes, the material strength is size-dependent due to strain gradients\cite{Doerner:1986aa,Nix:1998aa,Vlassak:1993aa,Begley:1998aa,Wei:2003aa,Hutchinson:2000aa,Uchic:2004aa,Uchic:2002aa,Fleck:2001aa,Aifantis:1984aa,Aifantis:1999aa} generated due to the absence of typical gradient-free dislocation motion and multiplication mechanisms. Furthermore, macroscopic work hardening~\cite{Anderson:2017aa, Devincre:2008kx} is replaced by a wealth of abrupt plastic events \cite{csikor2007dislocation,dimiduk2006scale,miguel2001intermittent,Maas:2013aa,koslowski2004avalanches} that originate in both the presence of dislocation correlations, as well as the dramatic small volume effect of mobile dislocations forming geometric steps on free pillar surfaces~\cite{miguel2001intermittent,dimiduk2006scale,sammonds2005deformation,Weinberger:2008aa,El-Awady:2011aa}. Abrupt plastic events are common in avalanche phenomena of various disordered non-equilibrium systems across length scales~\cite{uhl2015universal, fisher1997statistics, ben2008collective, demirel1996friction,papanikolaou2017avalanches}, especially elastic interface depinning phenomena, with which crystals share similar, but not identical, avalanche statistics~\cite{ispanovity2014avalanches}. However, in typical crystal ``depinning" modeling attempts~\cite{Zaiser:2005qf,Zaiser:2006ve,Dahmen:2009cr,Jagla:2010bs,Moretti:2004fv}, avalanches are caused by a direct competition of elastic loading and long-range elastic interactions with quenched disorder, without temporal bursts in the number of elastic degrees of freedom. In contrast, dislocations in nano-crystals can also nucleate, multiply and deposit on free boundaries~\cite{Ryu:2015aa,shan2008mechanical,Bulatov:2006yg}, naturally causing additional frustration that may influence the statistical avalanche behavior~\cite{papanikolaou2012quasi,Jagla:2010bs,Jagla:2010aa,Jagla:2007aa}. Here, in the context of an explicit dislocation dynamics model, we show that the competition between two different ways to mediate plastic slip -- dislocation nucleation and over-damped dislocation mobility (\ie dislocation drag) -- leads to a distinct rate effect on the avalanche statistics that becomes more pronounced for stress-controlled loading conditions. We interpret the phenomenon in terms of a spatial integration of avalanche behaviors across slip planes~\cite{papanikolaou2012quasi}. This is a generic mechanism in bifurcation processes such as the Frank-Read nucleation of a single dislocation, and thus we argue that the proposed effect should extend to 3D-DDD models~\cite{papanikolaou2017avalanches,cui2016controlling}.

Dislocation avalanches~\cite{Weiss:2000vn} have been observed  experimentally in diameter-$D$ micro and nano pillar compression studies~\cite{dimiduk2006scale, fressengeas2009dislocation,weiss2007evidence} where power law statistics for the sizes $S$ of the form $P(S)=S^{-\tau}\cP(S/S_0)$ has been established, where $\tau\in(1.2,1.8)$, $S_0\sim D$ and $\cP$ resembles an exponential cutoff function~\cite{friedman2012statistics}. Two~\cite{Ispanovity:2010aa,ispanovity2014avalanches,Miguel:2002aa,Miguel:2001aa,Moretti:2004aa,Zapperi:2001aa,Laurson:2006aa,Laurson:2010aa,Ovaska:2015aa,Groma:2012aa,Tsekenis:2011aa,Salman:2011fj,Salman:2012uq,Kale:2014nx} and three \cite{El-Awady:2011aa,Krebs:2017aa,Ispanovity:2010aa,Ispanovity:2013aa,Parthasarathy:2007aa,Rao:2008aa,Stricker:2015aa} dimensional models of atomic displacements or/and discrete dislocations simulations\cite{Miguel:2001aa,Ispanovity:2010aa,ispanovity2014avalanches,crosby2015origin,Swan:2018fk}  have established that $\tau\sim1.5$~\cite{Tsekenis:2013aa} or lower~\cite{Ovaska:2015aa}, regardless of loading protocols~\cite{csikor2007dislocation,uhl2015universal}, even if there are still  various issues on how the statistics is estimated~\cite{papanikolaou2017avalanches, Tsekenis:2011aa,ispanovity2014avalanches} However, recent 3D Discrete Dislocation Dynamics (3D-DDD) studies~\cite{cui2016controlling} showed that avalanche statistics strongly depend on the loading protocol, where power law statistics with $\tau\sim1.5$ only exhibited in stress-controlled (SC) loading. In addition, recent experiments and continuum modeling~\cite{Weiss:2015kl,papanikolaou2012quasi,Zhang:2017qa} have been suggesting that $\tau$ may take much larger values, with possible reasoning focused on internal disorder or/and thermal relaxation mechanisms such as cross-slip.

The effect of loading protocols on the statistical behavior of nanopillar compression response has been studied recently~\cite{maass2015slip, Sparks:2018cr}, even though the connection between stress rate $\dot{\sigma}$ (in SC) and strain rate $\dot{\epsilon}$ (in displacement-controlled loading (DC)) has been lacking at small rates. In contrast, at large loading rates ($>10^3/s$) and in the macroscale, it is well known that crystals exhibit a sharp increase of the flow stress due to viscoplastic {\it dislocation drag effects} when strain rate is higher than $\sim 5000/s$~\cite{armstrong2008high,murphy2010strength,tong1992pressure,Clifton:2000vf}. This fact has been well verified in DDD simulations~\cite{agnihotri2015rate,song2016discrete, hu2017strain} and originates in the natural competition between the timescale for a dislocation to move a Burgers vector distance at terminal speed and the timescale for dislocation ``nucleation" at a source (for example, at a pinned bulk segment -- Frank-Read source)~\cite{hirth1982theory}. How does this competition translate to the statistical behavior of plasticity avalanches in small volumes at rates smaller than $10^2/s$? 

In this paper, we consider a minimal model of crystal plasticity for uniaxial compression in small volumes. By ``minimal", we imply a model that respects: i) the energetics of room temperature crystal deformation being mediated by dislocations gliding along slip planes of at least one slip system ii) the fact that small-volume crystalline plastic deformation originates in nucleation, iii) open boundaries absorb dislocations. In order to maximize statistical sampling and computational efficiency, we perform simulations of 2D samples using a benchmarked dislocation dynamics model~\cite{papanikolaou2017obstacles,vandergiessen1995} that displays the basic phenomenology of nanocrystalline compression: Size effects in the material yield strength and emergent crackling noise.
For pure elasticity, SC and DC loading modes can be compared by using $\dot{\sigma}=E^{*}\dot{\epsilon}$, where $E^{*}=\frac{E}{1-\nu^{2}}$ is the equivalent modulus for plane strain applications and $\nu$ is the Poisson's ratio. The loading strain-rate $\dot{\epsilon}$ is varied from $10$/s to $10^{5}$/s. 
The model crystal  is initially stress and mobile-dislocation free. This is analogous to a well-annealed sample, yet with pinned dislocation segments that can act either as dislocation sources (\eg Frank Read sources) or as obstacles. Dislocations are generated from randomly distributed point sources when the resolved shear stress crosses a random threshold for a finite time $10ns$~\footnote{this process mimics the physical process of Frank Read source multiplication}. The nucleated dislocation pair is placed at a distance $L_{\rm nuc}=E/(4\pi(1-\nu^2))b/\tau_{\rm nuc}$  and for our system parameters, it is $35nm$ on average~\footnote{thus, the plastic strain generate by a nucleated dipole (which will be put at 35nm apart along a 60 degree slip plane) is $b*sin 60^{\circ}/h * 35nm*cos 60^{\circ}/w \simeq 10^{-6}$}.  Randomly distributed point obstacles account for precipitates and forest dislocations on out-of-plane slip systems. Microstructural parameters are chosen based on a previous study~\cite{papanikolaou2017obstacles} that matches various experimental facts. 

The timescale competition in this model is generic and present not only in all dislocation dynamics models, but also in generic non-equilibrium processes~\cite{Sahni:1983aa}. Its basic origin can be distilled by considering an imperfect pitchfork bifurcation: $d\eps/dt = \sigma + \mu \eps - \eps^3$, where $\eps, \sigma$ are scalars resembling strain and stress variables, and $\mu$ is a mobility parameter. Neglecting dislocation interactions, on a slip plane without sources but a mobile dislocation, $\mu=\mu_{\rm drift}$ is negative and the relaxation timescale for {\it every} incremental step of $\sigma$ is $\tau_{\rm drift}=|\mu_{\rm drift}|^{-1}$. However, if a dislocation source is present without any mobile dislocations, then $\mu=\mu_{\rm nuc}>0$ due to the existence of the two states with and without a dislocation pair, and the relaxation timescale during dislocation increments is typically $\tau_{nuc}=\mu_{nuc}^{-1}$. Typically, $\tau_{\rm nuc}\gg\tau_{\rm drift}$, so increments of $\sigma$ will typically be accomodated by nucleation events. However, if a system of such possible bifurcations interact (if multiple dislocation sources are present), then mutual dislocation interactions may cause a frustrating situation where the disparity of relaxation times may cause a complexity in the evolution dynamics. In our DDD model, the two timescales are concerned with the nucleation and propagation of single dislocations, where the timescale for a dislocation to move by a Burgers vector distance when the applied stress is near the dislocation nucleation stress $B/ \tau_{\rm nuc}\sim 2\times10^{-3}ns$ where $B$ is the linear drag coefficient.

Driven by local stress-induced forces~\cite{hirth1982theory}, dislocations follow athermal dynamics with mobility $\mu_d$. Sample lateral surfaces are free for dislocations to escape from the surfaces. Samples (aspect ratio $h/w=4$) are assumed to carry single slip plasticity oriented at $30^{\circ}$ (\cf Fig.~\ref{fig:1}(a)). Dislocation sources (red dots) and obstacles (blue dots) are located on slip planes, spaced 10$b$ apart, with $b=0.25$nm the Burgers vector's length. The Young's modulus is assumed $E=70$ GPa and $\nu=0.33$. As it may be seen in Fig.~\ref{fig:1}(b), a significant difference between two loading rates (SC) can be seen through strain patterning at the same final strain (5\%): plasticity is localized (Fig.~\ref{fig:1}(c)) for small loading rates while it is relatively uniform for a high loading rate (Fig.~\ref{fig:1}(d)).

\begin{figure}[t]
\includegraphics[width=0.48\textwidth]{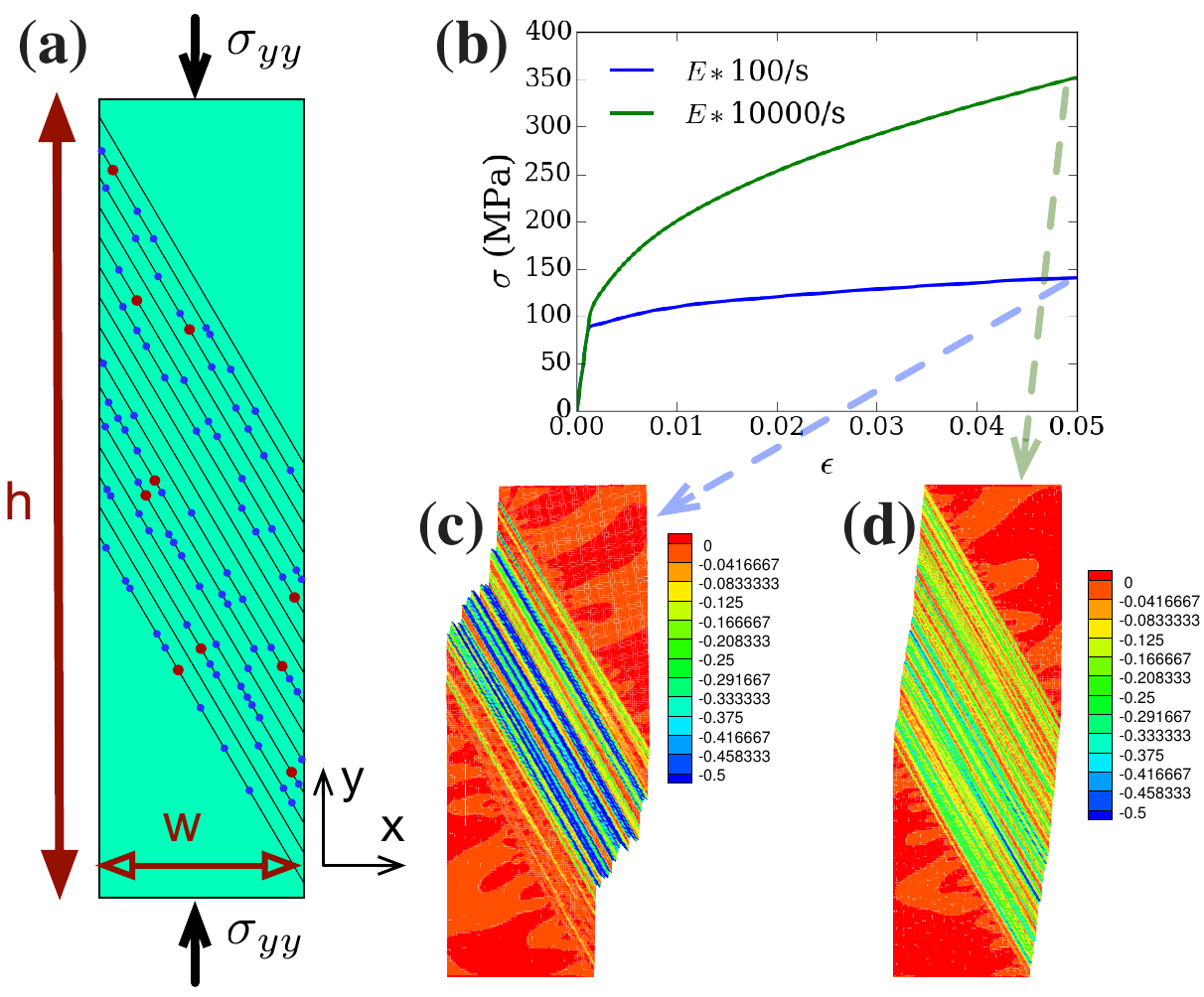}
\caption{{\bf The model. }{\bf (a)} The pillar has width $w$ and aspect ratio $h/w$=4. Single slip system which oriented at $30^{\circ}$ relative to y axis is used. Distance between planes is 10$b$ where $b=0.25$nm is the magnitude of the Burgers vector. Red dots stand for dislocation sources while blue dots represent dislocation obstacles. {\bf (b)} Sample stress strain curves of compression at high ($10^5/$s) and low ($10^2/$s) stress rates $\dot{\sigma}$. {\bf (c)} strain pattern after deformation at low $\dot{\sigma}$,  {\bf (d)} strain pattern after deformation at high $\dot{\sigma}$. }
 \label{fig:1}
\end{figure}

\begin{figure}[t]
\includegraphics[width=0.48\textwidth]{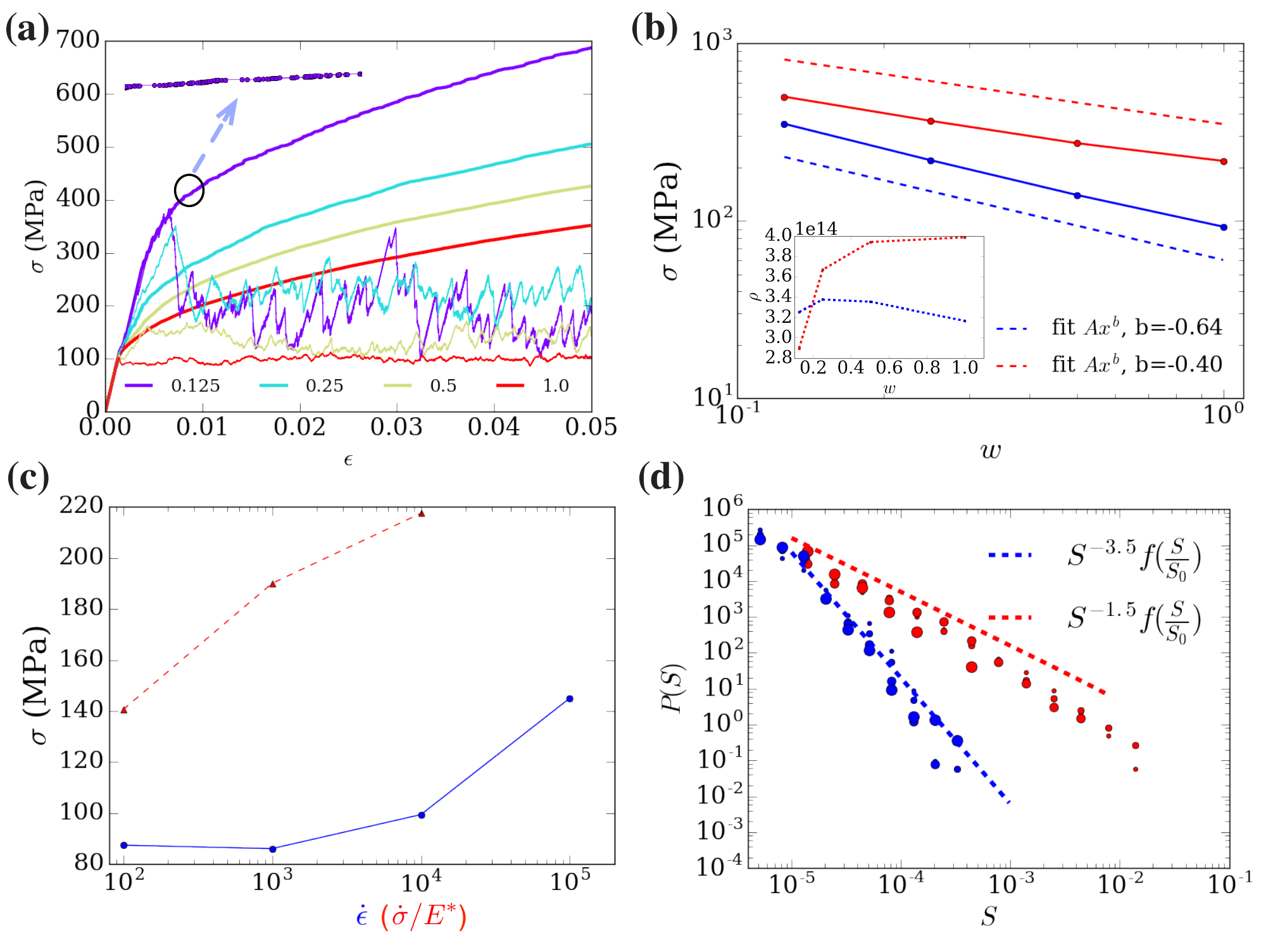}
\caption{ {\bf Effect of loading protocol: Stress-Controlled (SC) vs. Displacement-Controlled (DC).}
{\bf (a)} Stress-strain curves of different $w$ using two different loading protocols. The strain rate $\dot{\epsilon}$ is $10^4$/s in DC and stress rate $\dot{\sigma}$ is $E^{*}*10^4$/s. A particular strain burst is shown; {\bf (b)} Size effect of flow stress at 2\% strain (blue stands for DC and red stand for SC, results are based on 50 realizations). The inset shows the dependence of dislocation density on $w$ at 2\% strain for different loading protocols. {\bf (c)} Dependence of flow stress (for $w=1\mu\rm{m}$) on rate. Strain rate $\dot{\epsilon}$ is used in DC (blue curve) while the elastic corresponding stress rate $\dot{\sigma}=E^{*}\dot{\epsilon}$ is used in SC (red curve). {\bf (d)}: Events (strain jumps) statistics for different loading protocols, different point size represents different $w$. blue: DC, red: SC. Strain jump in DC mode is calculated according to the method in~\cite{cui2016controlling}.}
 \label{fig:2}
\end{figure}

As expected and shown in Fig.~\ref{fig:2}~(a), SC  leads to hardening while DC to softening, with the discrepancy becoming dramatic as system size decreases to sub-micron dimensions. Typical size effects ($\sigma_Y\sim w^{-0.4-0.6}$) are seen for both loading protocols~(\cf Fig.~\ref{fig:2}(b)), despite the fact that dislocation density at 2\% strain, shown in the inset, increases with increasing $w$ in different ways depending on the loading conditions. In addition, the flow stress shows a rate dependence for both loading conditions (see Fig.~\ref{fig:2}(c)), even though  DC shows weaker dependence. The dislocation density and flow stress dependences on the rate suggest that SC rates {\it statistically resemble} larger DC rates. This conclusion is also supplemented by  avalanche statistics (\cf Fig.~\ref{fig:2}(d)):
}{ In SC, event size is defined as $S=\sum_{i\; \in\; \left\lbrace\delta\epsilon^{i}>\epsilon_{\rm{threshold}}\right\rbrace}\delta \epsilon^i$; in DC, an event is characterized by stress drops $\delta\sigma$ which lead to temporary displacement overshoots -- thus, in order to compare the two loading conditions, a DC strain burst event size is defined as $S=\sum_{i\; \in\;\left\lbrace {-\delta \sigma^{i}}>\sigma_{\rm{threshold}} \right\rbrace}\delta \epsilon^i$~\cite{cui2016controlling}.

In this model, dislocation plasticity is loading rate dependent as there are two intrinsic time scales~\cite{agnihotri2015rate}: First, the dislocation nucleation time $t_{\rm{nuc}}$, which is chosen as $10~ns$ and can be associated to the dislocation multiplication timescale in other models. Second, the ratio between dislocation mobility and material Young's modulus $B/E$ which is chosen as $10^{-6}$ ns. These parameters are consistent with recent single-crystal thin film experiments~\cite{xiang2006bauschinger,nicola2006plastic}.
Phenomenology in metallurgy~\cite{follansbee1988constitutive,tong1992pressure,clifton1990high},  suggests that at low rates the flow stress is controlled by dislocation nucleation while above a certain strain rate ($\sim1000-5000/$s), it is mainly controlled by dislocation drag. Fig.~\ref{fig:2}(c) shows the rate effect under SC and DC conditions. For DC and at strain rates higher than 5000/s, there is a strong flow stress rate dependence. In SC, the drag regime starts when stress rate is $E^{*}*10^2$/s.  The origin of this strain-rate crossover is clear in this model: It is clear that during the nucleation events, strain-increments are necessarily mediated by dislocation drag; for mobile dislocation density $\rho\simeq10^{12}/m^2$ and flow stress $\tau_f\simeq 50MPa$, the strain generated by moving dislocations in time-intervals of nucleation time can be up to $0.5\rho\tau_f b^2 / B * t_{\rm nuc} \simeq \times 10^{-4}$. Thus, for strain-rates greater than $10^2/s$, the strain increment required during $\tau_{\rm nuc}$ is $\dot\eps \times \tau_{\rm nuc} > 10^{-6}$, which implies that the nucleation-induced strain is not adequate. Thus, it is plausible that for $\dot\eps>10^{-3}/s$, dislocation drag takes over the dynamics of dislocations.

While both DC and SC display a flow stress rate effect, their statistical noise behavior is very different: As shown in  Fig.~\ref{fig:2}(d), the plastic events statistics based on stress strain curves shown in Fig.~\ref{fig:2}(a), have different $\tau$ exponents:  While plastic events show power law behavior, $\tau$ is close to $3.5$ for DC and 1.5 for SC. 

\begin{figure}[t]
\includegraphics[width=0.5\textwidth]{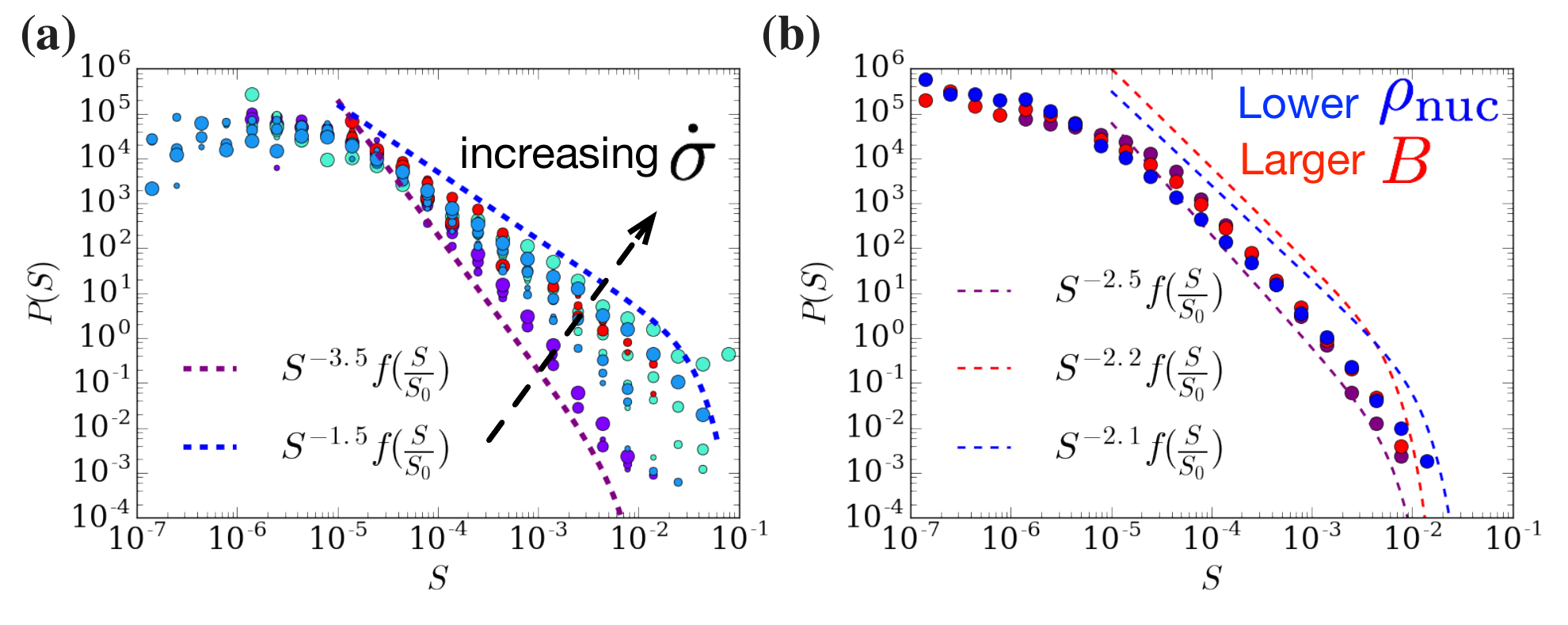}
\caption{ {\bf SC Rate Effect Crossover.~}
{\bf(a):} Event statistics for different $\dot{\sigma}$ using SC. The effective $\tau$ changes from $\sim-3.5$ for $\dot{\sigma}=E^{*}*10$/s to $\sim-1.5$ for $\dot{\sigma}=E^{*}*10^{4}$/s. 
{\bf (b):} Effect of dislocation source density $\rho_{\rm{nuc}}$ and mobility $B$ on power law exponent: when $\dot{\sigma}=E^{*}*10^{2}$/s, changing $\rho_{\rm{nuc}}$ from 60$\mu\rm{m}^{2}$ (purple curve) to 15$\mu\rm{m}^{2}$ (blue curve) leads to the exponent changing from -2.5 to -2.1. Increasing $B$ from $10^{-4}$ Pa.s to $10^{-3}$ Pa.s results in the change of exponent from -2.5 to -2.2.}
 \label{fig:3}
\end{figure}

The avalanche size distribution exponent discrepancy between SC and DC disappears at high stress loading rates: Fig.~\ref{fig:3}(a) shows avalanche statistics for different stress rate which varies from $\dot{\sigma}=E^{*}*10$/s to $\dot{\sigma}=E^{*}*10^{4}$/s. Power law events distribution appear for all stress rates, yet with different exponent which changes from -3.5 for $\dot{\sigma}=E^{*}*10$/s to -1.5 for $\dot{\sigma}=E^{*}*10^{4}$/s. The dependence of the exponent on the stress rate clearly indicates that there is an intrinsic connection between event statistics and dislocation drag. In order to verify the connection, in Fig.~\ref{fig:3}(b) red curve, we increase the dislocation mobility $B$ by which the drag effect is enhanced, it is seen that the exponent changes from -2.5 to -2.2. Dislocation drag effect will also magnify when dislocation nucleation effect is weakened due possibly to dislocation cross-slip and other mechanisms (since the main source of plasticity will be the moving of dislocations instead of nucleations of new dislocations). This can be seen in Fig.~\ref{fig:3}(b) blue curve, when lower dislocation source density is used, the exponent changes from -2.5 to -2.1. 

\begin{figure}[t]
\includegraphics[width=0.48\textwidth]{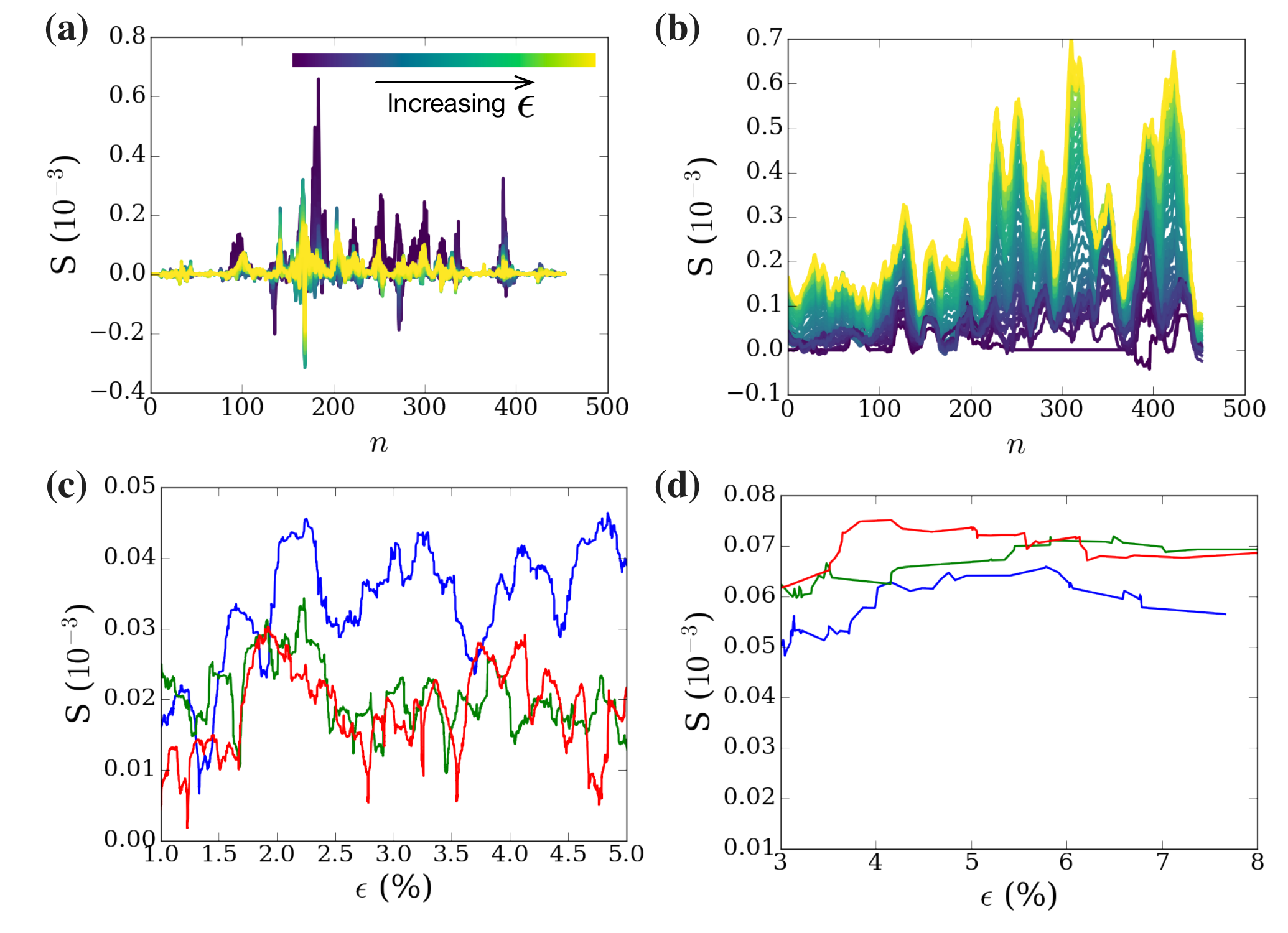}
\caption{ {\bf Spatial and temporal event distribution in SC.}
 Event distribution on all slip planes during the loading up to 10\% strain for small $\dot{\sigma}$ ({\bf (a)}) and for large $\dot{\sigma}$ ({\bf (b)}): $n$ is the total number of slip planes in the model. The color changes from dark purple to yellow with increasing loading strain.  
{\bf (c)}: Average avalanche size for small $\dot{\sigma}$ in a single sample.
{\bf (d)}: Average avalanche size for large $\dot{\sigma}$ in a single sample.}
 \label{fig:4}
\end{figure}
Power law avalanche behavior in the elastic response of disordered systems has been well established~\cite{Zaiser:2006ve,Zaiser:2005qf,fisher1997statistics,dahmen2011simple,Dahmen:2009cr}. In the context of nanopillars, the dislocation ensemble should be the homogeneously disordered elastic system and in this case, the spatial distribution of events on all slip planes should be on average flat or display relatively small fractal exponents~\cite{Fisher:1998aa} in the absence of localization. However, crystal plasticity is known to be unstable to strain localization~\cite{Asaro:2006aa}. In Fig.~\ref{fig:4}(a) and (b), we plot events spacial distribution along all slip planes for the whole loading process (from small $\epsilon$ to large $\epsilon$ which is represented by the color map from purple to yellow). Fig.~\ref{fig:4}(a) shows the event spatial distribution for a smaller loading rate. It can be seen that events are localized around certain slip planes, moreover, events do not always happen at the same slip planes. By contrast, the event distribution shown in Fig.~\ref{fig:4}(b) for higher loading rate is more uniform among slip planes, furthermore, events always happen at the same active slip planes which is similar to having an propagating interface. Additionally, we plot event size with increasing strain (S vs. time). Very clear oscillatory-like behaviour emerges for small stress rate shown in Fig.~\ref{fig:4}(c) while no periodicity is observed for higher stress rate. These results are strikingly similar to the avalanche oscillator found in~\cite{papanikolaou2012quasi}. 

The onset of quasi-periodic response at small rates, in the absence of overall weakening in this model, is the outcome of the interplay between a timescale competition (as in other elasticity models~\cite{Jagla:2010bs}) and a distinct feature of small volumes: \ie Free boundaries that may absorb propagating dislocations. Due to this property, it is natural to expect an integration of avalanche behaviors, dependent on the resetting behavior that emerges from absorption and re-nucleation of dislocations at various slip planes. The overall effect can be thought of as originating within a relaxation process (nucleation) that contributes to slip, in addition to mobile dislocation motion. This is the type of coarse-grained dislocation modeling that was pursued in Ref.~\cite{papanikolaou2012quasi} and its analysis leads to critical power law exponents that are higher than typical ones ($\sim3/2$). Local heterogeneity biases the integration of the size probability distribution of the conventional depinning models. In this paper, through dislocation dynamics simulations, we connect plasticity local heterogeneity to strain rate effect: the lower loading rate results in the higher heterogeneity which leads to a higher power law exponent. If $P(S,k)\sim S^{-\tau_0}e^{-k S}$, with $k$ a cutoff parameter then this spatiotemporal integration leads to an effective probability distribution:
\bea
P_{int}(S)=\int_{0}^{\infty} g(k') P(S,k') dk'
\eea
where $g(k')$ is the weight factor that characterizes the contribution of various sub-critical, quasi-localized spatial contributions to slip events and depends on the applied loading rate. This weight factor $g(k')$ contains a natural $k'\rw0$ limit, due to the quasi-periodic resetting, which in many cases takes the form of a power-law~\cite{papanikolaou2012quasi}, thus identifying a novel exponent $g(k')\sim k'^{\alpha}$. Thus, for the critical aspect of $P_{int}(S)\sim S^{-\tau_0 - \alpha - 1}$, with the ultimate avalanche size exponent being,
\bea
\tau= \tau_0 + \alpha + 1
\eea
For the current model, by the analysis of Figs.~\ref{fig:4}(a, b), we can estimate $\alpha$: If we assume that each 3 nearby slip planes are locally independent from the rest of the system, then the max event size in that area can provide an estimate of the cutoff scale ($k_0\sim 1/S_0$). Then, the distribution of $k_0$'s provides the exponent. We find that $\alpha\simeq 1$ by plotting the histogram of events that  considering $\tau_0\simeq1.5$. However, the statistics has not been exhaustive enough to justify a precise identification of these exponents. 

In conclusion,  we provided strong evidence through an explicit model of crystal plasticity for nanopillar compression, that the statistical behavior of nanocrystal plasticity forms a novel universality class that is distinct from other plasticity behaviors such as amorphous BMGs and granular systems~\cite{papanikolaou2017avalanches}. We find that the free nanoscale boundaries and the competition between dislocation nucleation and drag conspire to cause the emergence of unconventional quasi-periodic avalanche bursts and higher critical exponents as  strain rate decreases. While the investigated strain-rates and the associated transition emerges at relatively high loading rates, the experimentally relevant quasi-static response may be controlled by the same qualitative behavior~\cite{Sparks:2018cr}, or more timescales might be in competition. Plasticity is locally heterogeneous, both spatially and temporally, and this reason lies behind the rate dependence of the avalanche distribution exponent.

\begin{acknowledgments}
We would like to thank I. Groma, P. Ispanovity, R. Maass, L. Ponson for encouraging and insightful comments. This work is supported through awards DOC - No. 1007294R (SP) and DOE-BES DE-SC0014109. This work benefited greatly from the facilities and staff of the Super Computing System (Spruce Knob) at West Virginia University.
\end{acknowledgments}

}


\end{document}